%% file: main.tex
\documentclass[runningheads]{llncs}
\usepackage{graphicx}
\usepackage{booktabs}
\usepackage{multirow}
\usepackage{tikz}
\usepackage{tikz-qtree}
\usepackage{pgfplots}
\usetikzlibrary{patterns}
\usepackage{subfigure}
\usepackage{url}
\usetikzlibrary{matrix}
\usepgfplotslibrary{groupplots}
\pgfplotsset{compat=newest}
\usepackage{tablefootnote}
\usepackage{comment}
\usepackage{makecell}
\usepackage{enumitem}
\usepackage{booktabs}
\usepackage{amsmath,amssymb}
\usepackage{bbding}
\usepackage[ruled,vlined,linesnumbered]{algorithm2e}
\usepackage{multirow}
\usepackage[justification=centering]{caption}

\usepackage[T1]{fontenc}
\usepackage{graphicx}
\begin{document}
\title{Meta-learning Enhanced Next POI Recommendation by Leveraging Check-ins from Auxiliary Cities
%\thanks{Supported by organization x.}%
}

\author{Jinze Wang\inst{1}\and
Lu Zhang\inst{2}{\textsuperscript(\Envelope \textsuperscript)}\and
Zhu Sun\inst{3,4}\and
Yew-Soon Ong\inst{3,5}}
\authorrunning{J. Wang et al.}
\institute{Macquarie University, Balaclava Rd, Macquarie Park NSW 2109\\
%\email{oliver.jinzewang@gmail.com}\\ 
\and
Chengdu University of Information Technology, Chengdu, China\\
\and
Centre for Frontier AI Research, A*STAR, Singapore\\
\and
Institute of High Performance Computing, A*STAR, Singapore\\
%\email{sunzhuntu@gmail.com }
\and
School of Computer Science and Engineering, Nanyang Technological University, Singapore\\
\email{zhang\_lu010@outlook.com}\\
}
\maketitle              % typeset the header of the contribution
\begin{abstract}
Most existing point-of-interest (POI) recommenders aim to capture user preference by employing city-level user historical check-ins, thus facilitating users' exploration of the city. However, the scarcity of city-level user check-ins brings a significant challenge to user preference learning. Although prior studies attempt to mitigate this challenge by exploiting various context information, e.g., spatio-temporal information, they ignore to transfer the knowledge (i.e., common behavioral pattern) from other relevant cities (i.e., auxiliary cities). In this paper, we investigate the effect of knowledge distilled from auxiliary cities and thus propose a novel Meta-learning Enhanced next POI Recommendation framework (MERec). The MERec leverages the correlation of check-in behaviors among various cities into the meta-learning paradigm to help infer user preference in the target city, by holding the principle of ``paying more attention to more correlated knowledge''. Particularly, a city-level correlation strategy is devised to attentively capture common patterns among cities, so as to transfer more relevant knowledge from more correlated cities. Extensive experiments verify the superiority of the proposed MERec against state-of-the-art algorithms.

\end{abstract}
\input{section/intro_v1.tex}

\input{section/relatedwork_v1.tex}
\input{section/datacollection_v1.tex}

\input{section/framework_v1.tex}

\input{section/experiments_v1.tex}

\input{section/conclusion}

% \subsubsection{Acknowledgements} Please place your acknowledgments at
% the end of the paper, preceded by an unnumbered run-in heading (i.e.
% 3rd-level heading).


\bibliographystyle{splncs04}
\bibliography{mybibliography}

\end{document}

%% file: section/intro_v1.tex
\section{Introduction}
Next POI recommendation, which aims to recommend POIs for users that they are most likely to visit in the future, benefits both location-based social network services, e.g., Foursquare (\url{foursquare.com}), and individuals.
As users' activities typically limit within a city, most existing studies exploit the city-level user check-in records to develop next POI recommenders. 
Table~\ref{tab1} shows the statistics of user-POI interactions for four cities on Foursquare, which are widely explored in prior studies\textcolor{black}{~\cite{zhang2021interactive,zhangnext}}. 
We can observe that CAL with relatively higher density being 1.06\%, while the extremely lower density is 0.05\% in NYC. Obviously, the sparsity of user-POI interactions in many cities severely hinders the capability of existing approaches for more accurate user preference learning.

To ease this issue, various context information, e.g., spatial and temporal contexts, has been widely exploited in existing next POI recommenders.
Specifically, most current research devotes to capturing the spatio-temporal relations between users and POIs. They are built upon various techniques, ranging from matrix factorization~\cite{lian2014geomf,wang2021footprint}, Markov chain models~\cite{cheng2013you}, to advanced deep learning frameworks, e.g., recurrent neural networks~\cite{zhang2021interactive} and graph neural networks~\cite{qian2019spatiotemporal}. 
However, they are restricted by insufficient training data for more accurate user preference learning due to the sparse user-POI interactions within a city. 

Intuitively, users' check-in behaviors among different cities may share common patterns. This motivates us to conduct an in-depth analysis of the check-in records across different cities (i.e., auxiliary cities), and transfer useful knowledge from such cities for assisting user preference inference within the target city.
However, non-overlapping visited POIs between different cities bring challenges in knowledge transfer, that is, blindly leveraging check-in behaviors from auxiliary cities to augment the target city may result in harmful knowledge transfer. We thus seek to investigate two fundamental problems when transferring knowledge from auxiliary cities to the target city as follows. 

\input{tables/statistics_of_datasets.tex}

% transfer categorical patterns
\textbf{(1)} \textit{\textbf{What to transfer?}} 
In e-commerce, overlapping items can be found on shopping sites in different regions. 
While in the city-level location recommendation scenario, non-overlapping visited POIs across different cities present a challenge to transferring common behavioral knowledge.
Fortunately, mining users' check-in behavioral knowledge over the categorical context (i.e., common category-level patterns) helps address this challenge.  For example, \textcolor{black}{the category transition \textit{Shop}\&\textit{Service}$\rightarrow$\textit{Food} are common to all four cities, which indicates that users in different cities are most likely heading to a restaurant after shopping.} By contrast, the transition \textit{Travel}\&\textit{Transport}$\rightarrow$\textit{Shop} is quite common only in SIN due to the developed public transportation. 
\textbf{(2)} \textit{\textbf{How to transfer?}} 
Although the common category-level patterns captured from auxiliary cities may enhance the recommendation quality for the target city, this inevitably introduces noise if we ignore the cultural diversity and geographical property of such cities. 
Hence, determining what extent we can transfer knowledge from the auxiliary cities to the target city is of great significance.

Accordingly, we propose a novel Meta-learning Enhanced next POI Recommendation (MERec) framework, which delicately considers the correlation of category-level behavioral patterns among different cities into the meta-learning paradigm, that is, paying more attention to more correlated knowledge. 
{Specifically, MERec mainly consists of two components: a \textit{two-channel encoder} to capture the transition patterns of categories and POIs, whereby a city-correlation based strategy is devised to attentively capture common knowledge (i.e., patterns) from auxiliary cities via the meta-learning paradigm;  
and a \textit{city-specific decoder} to aggregate the latent representations of the two channels to perform the next POI prediction on the target city.} 

Overall, our main contributions lie in three folds: (1) we are the first to study to what extent we can transfer knowledge from auxiliary cities to the target city via differentiating the correlation of category-level behavioral patterns; (2) we propose a novel meta-learning based framework -- MERec, which exploits both the transferred knowledge and user behavioral contexts within the target city to alleviate the data sparsity issue; and (3) we conduct extensive experiments on four datasets to validate the superiority of MERec against state-of-the-arts.

%% file: tables/statistics_of_datasets.tex
\begin{table}[t]
\centering
\scriptsize
\addtolength{\tabcolsep}{4pt}
\caption{\footnotesize Statistics of four datasets from Foursquare.}\label{tab1}
\begin{tabular}{c|c|c|c|c|c}
\toprule
& \#Users & \#POIs & \#Check-ins & \#Categories &Density \\ \midrule
Calgary (CAL)   & 435    & 3,013  & 13,911      & 293  &\textbf{1.06\%}     \\
Phoenix (PHO)   & 2,945   & 7,247  & 47,980      & 344 &\textbf{0.22\%}       \\
Singapore (SIN) & 8,648   & 33,712 & 355,337     & 398 & \textbf{0.12\%}     \\ 
New York (NYC)  & 16,387  & 56,252 & 511,431     & 420 & \textbf{0.05\%}      \\\bottomrule
\end{tabular}
\vspace{-0.25in}
\end{table}

%% file: section/relatedwork_v1.tex
\section{Related Work}\label{Sec: RW}

\noindent\textbf{Next POI Recommendation.} 
It predicts future POI visits for users based on their historical successive check-in behaviors. 
Early studies generally employ the property of Markov chain to model the sequential influence~\cite{cheng2013you,ye2013s,feng2015personalized}. Recently, recurrent neural network (RNN) based methods show great capability in capturing long-term sequential dependencies. Existing studies based on RNN and its variants mainly tend to exploit users' sequential check-ins by incorporating various context information, such as ST-RNN~\cite{liu2016predicting}, SERM~\cite{yao2017serm}, MCARNN~\cite{liao2018predicting} ATST-LSTM~\cite{huang2019attention}, and iMTL~\cite{zhang2021interactive}.
Despite the great success of these methods, 
most of them suffer from the issue of insufficient user check-ins in many cities, which heavily limits their performance improvements. In this sense, transferring knowledge from auxiliary cities to the target city brings the possibility to further enhance the user preference learning for the next POI recommendation.

\smallskip\noindent\textbf{Meta-learning for Next POI Recommendation.} 
Transfer learning (TL) aims to transfer knowledge from source domains to the target domain, which has shown strong capability in resolving the sparsity issue.
Existing TL-based approach~\cite{ding2019learning} focuses on the cross-city POI recommendation task due to the lack of large amount of overlapping user-POI interactions across cities. 
Meta-learning (ML) is able to transfer the knowledge learned from multiple tasks to a new task and has been recently introduced in next POI recommendation. 
For example, Chen et al.~\cite{chen2021curriculum} proposed CHAML by fusing hard sample mining and curriculum learning into a meta-learning framework. 
Sun et al.~\cite{sun2021mfnp} devised MFNP to integrate user preference and region-dependent crowd preference tasks in a meta-learning paradigm. 
Cui et al.~\cite{cui2021sequential} designed Meta-SKR by using sequential, spatiotemporal, and social knowledge to recommend next POIs. Meanwhile, Tan et al.~\cite{tan2021meta} developed the METAODE which models city-irrelevant and -specified information separately to achieve city-wide next POI recommendation. 
However, the aforementioned ML-based next POI recommenders ignore to attend the correlation of user behavioral patterns when transferring knowledge from auxiliary cities to the target city, i.e., paying more attention to more correlated knowledge.

%%%%%%%%%%

%% file: section/datacollection_v1.tex
\section{Data Analysis}\label{Sec: DC}
%% necessary for data analysis
There is a great necessity to analyze the correlation among different cities w.r.t. user check-in behaviors (see Table~\ref{tab1}), so as to better guide the knowledge transfer from auxiliary cities to the target city. 
It is, however, non-trivial due to the non-overlapping visited POIs across cities. 
Fortunately, POIs in various cities share the same categories, which inspires us to study the POI distribution and user behavioral patterns at the category level to uncover the correlation among cities. 

\begin{figure}[t] 
\centering 
\includegraphics[scale=0.2]{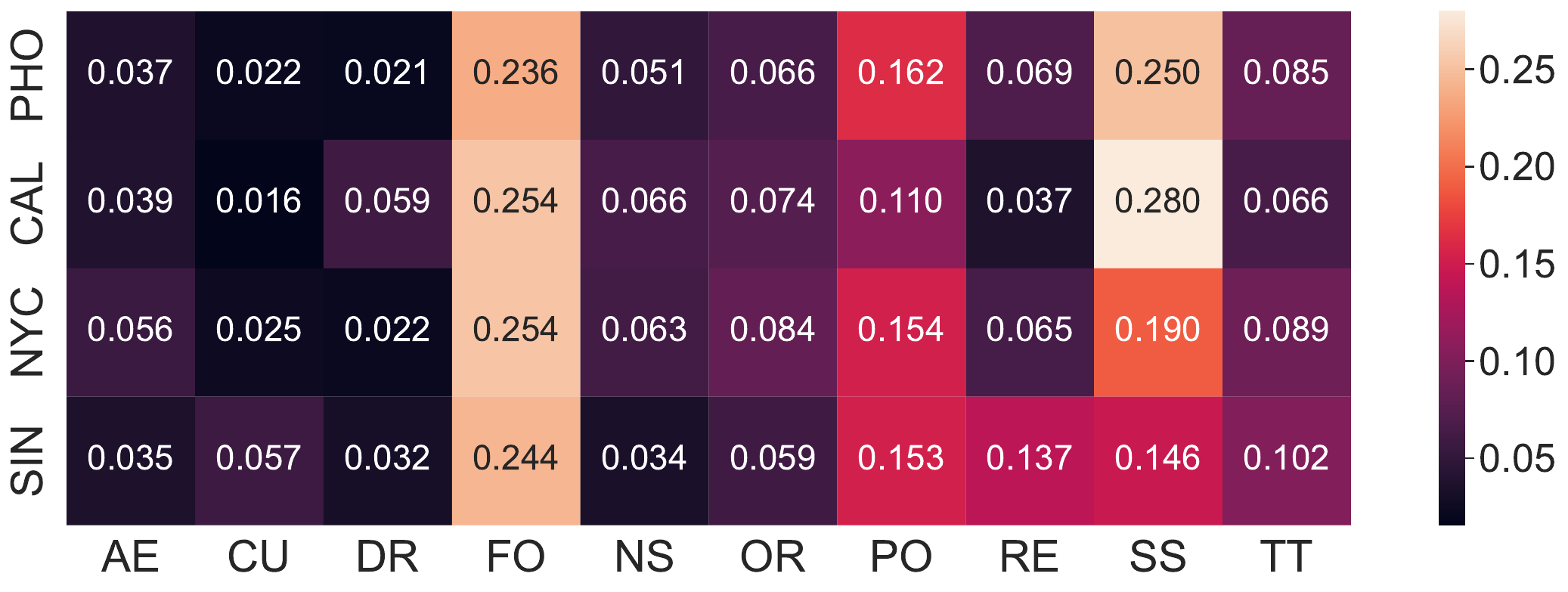} 
\vspace{-0.1in}
\caption{The distribution of POIs at category level among four cities.} 
\vspace{-0.25in}
\label{Fig:pec} 
\end{figure}

\smallskip\noindent\textbf{POI Distribution at  Category Level}. The number of POIs under each category varies a lot across cities due to different cultures and geography. Hence, we first study the nature of POI distributions among four cities to help explore the correlation of user behavioral patterns.
Specifically, all POIs are characterized by ten first-level categories~\cite{sun2021point}, including Arts \& Entertainment (AE), College \& University (CU), Drink (DR), Food (FO), Nightlife Spot (NS), Outdoor \& Recreation (OR), Professional \& Other Places (PO), Residence (RE), Shop \& Service (SS), and Travel \& Transport (TT). 
Fig.~\ref{Fig:pec} depicts the POI distribution at category level, where we note that cities exhibit high similarity in some categories while show dissimilarity in others.  
For example, the proportion of POIs under FO is relatively higher across the four cities, whereas the proportion of POIs under, e.g., AE, is lower than POIs under FO and SS. 
On the other hand, different cities show their unique characteristics, such as the higher proportion of CU-related POIs in SIN and the higher proportion of AE-related POIs in NYC.

\begin{figure}[t]
\centering
\subfigure[]{
\begin{minipage}[t]{0.24\linewidth}
\centering
\includegraphics[width=1.25in]{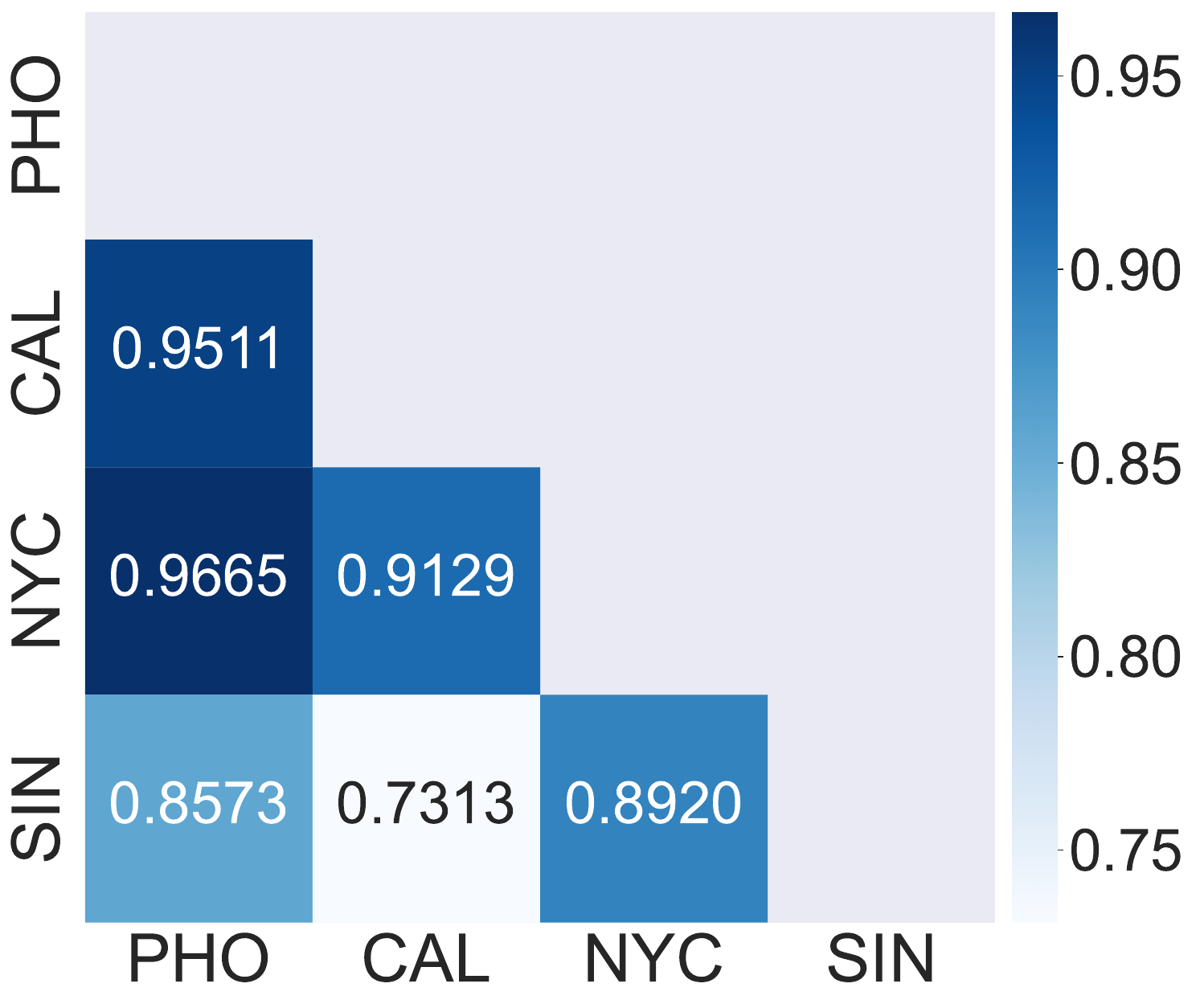}\label{fig:cor_city_POI_dist}
\end{minipage}
}
\subfigure[]{
\begin{minipage}[t]{0.24\linewidth}
\centering
\includegraphics[width=1.25in]{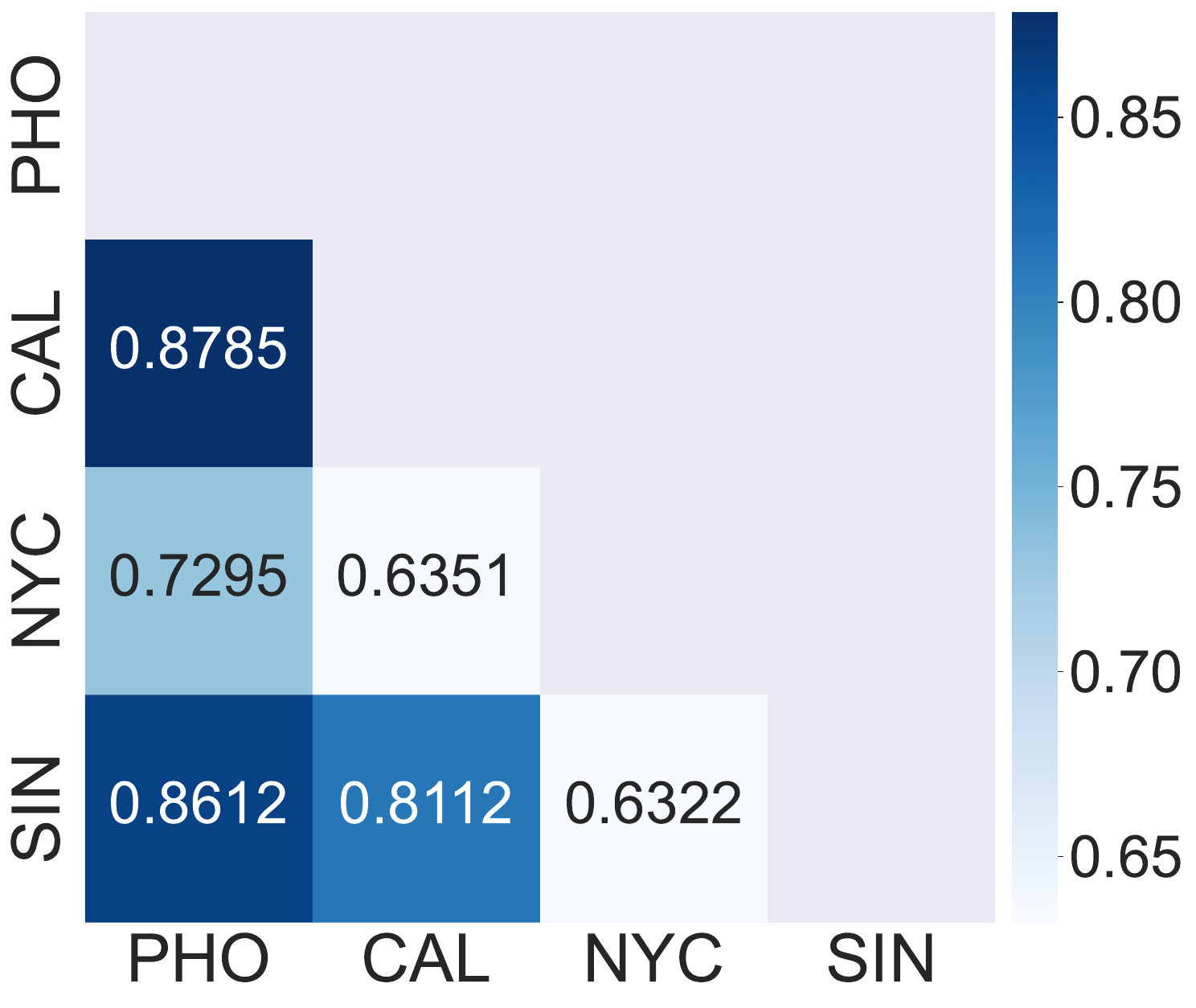}\label{fig:cor_city_behav}
%\caption{fig2}
\end{minipage}
}
\subfigure[]{
\begin{minipage}[t]{0.20\linewidth}
\centering
\includegraphics[width=1.1in]{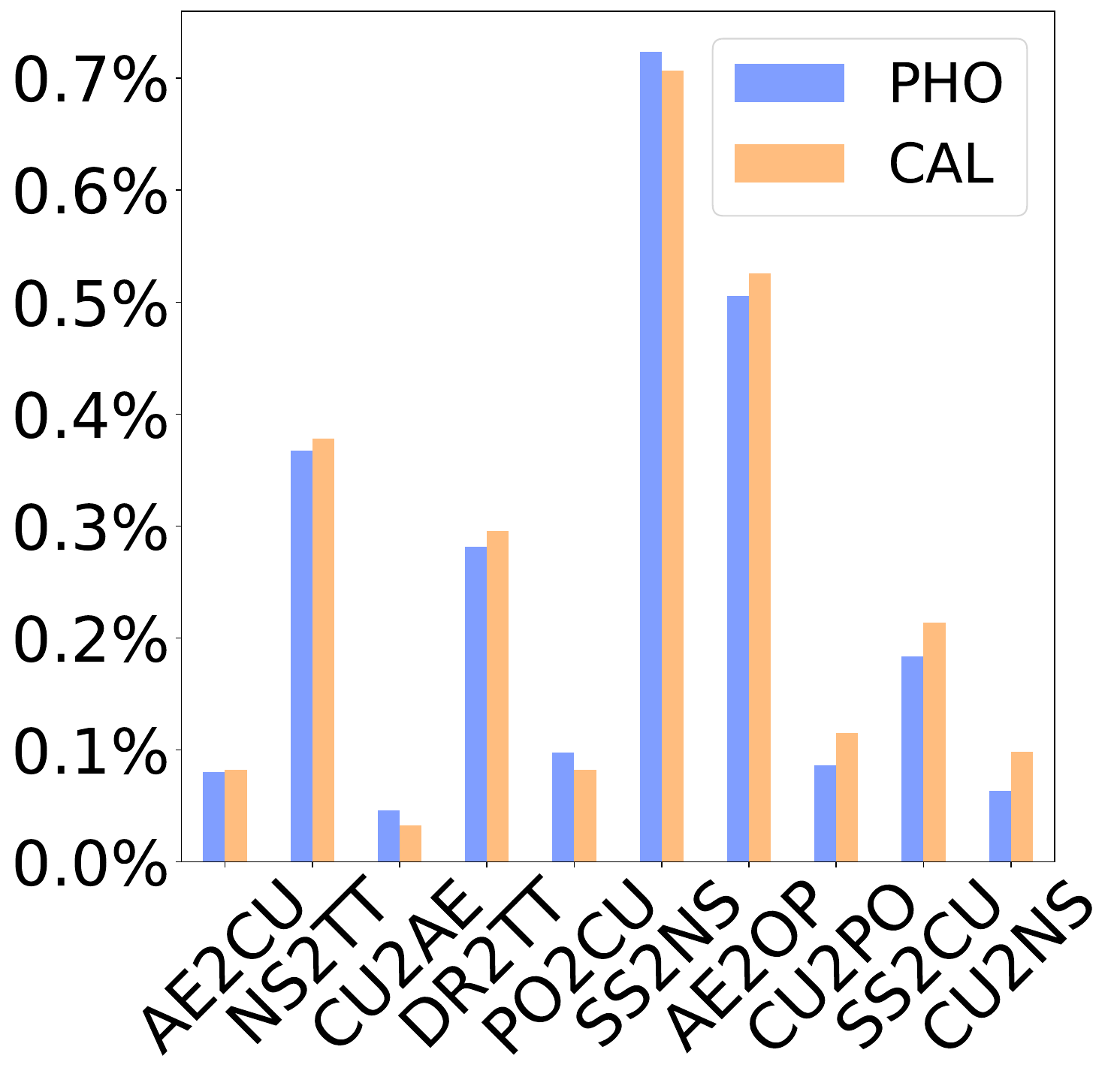}\label{fig:corr_PHO_CAL}
%\caption{fig2}
\end{minipage}
}
\subfigure[]{
\begin{minipage}[t]{0.22\linewidth}
\centering
\includegraphics[width=1.1in]{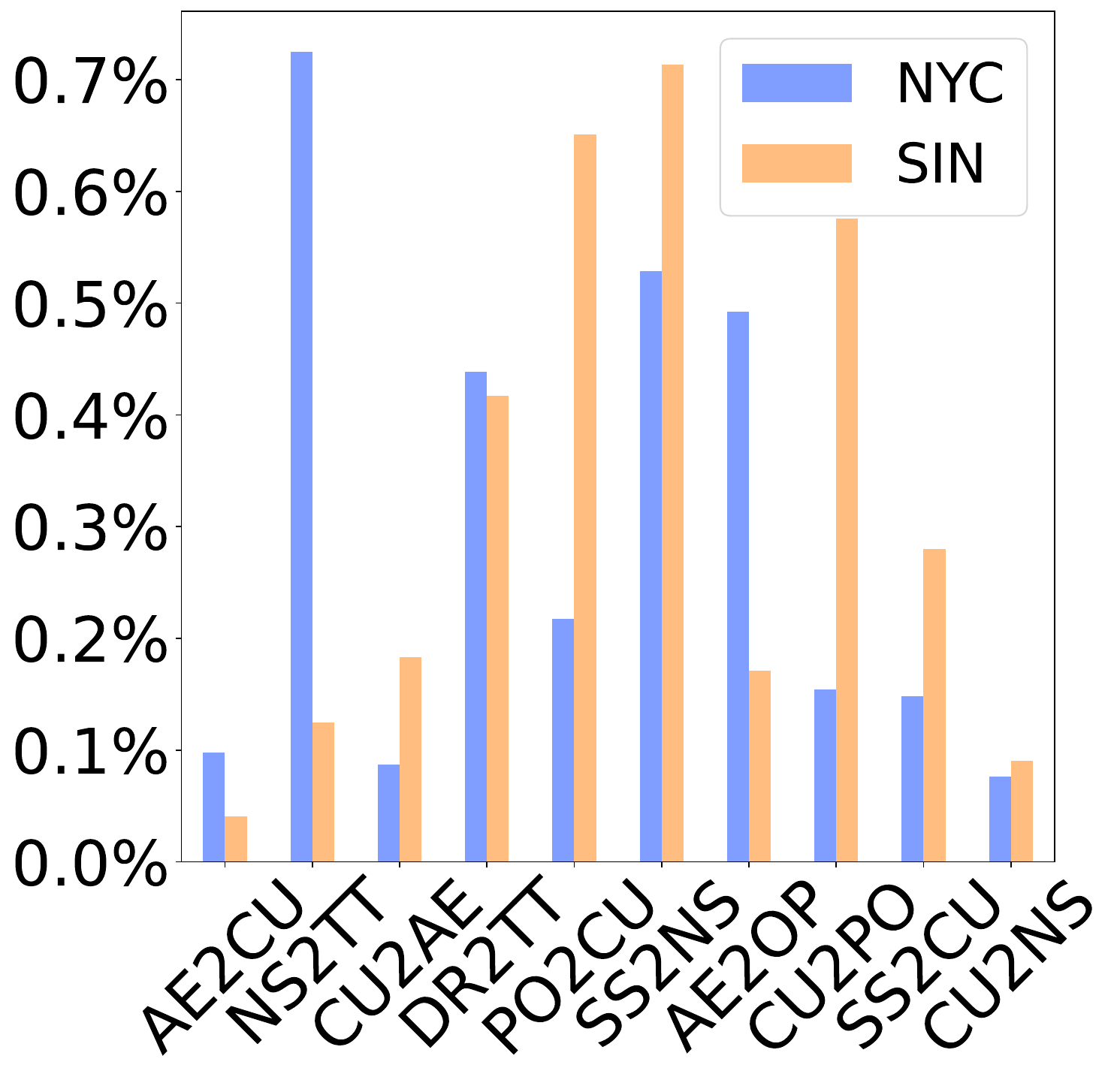}\label{fig:corr_NYC_SIN}
%\caption{fig2}
\end{minipage}
}
\centering
\vspace{-0.2in}
\caption{ (a-b) the correlation of four cities w.r.t POI distribution and behavioral patterns at category level; (c-d) two most correlated and least correlated cities.}
\vspace{-0.25in}
\end{figure}

\smallskip\noindent\textbf{Correlation of Cities w.r.t POI Distribution}. 
The POI distribution of each city enables us to further explore the correlation between cities, i.e., measuring the similarity of cities from the aspect of POI distribution. 
Specifically, given any two cities, $A^{poi} = [A_{1}^{poi}, A_{2}^{poi}\cdots A_{\mathcal{|C|}}^{poi}]$ and $B^{poi} = [B_{1}^{poi}, B_{2}^{poi}\cdots B_{\mathcal{|C|}}^{poi}]$ denote the POI distributions among $\mathcal{|C|}$ categories within city A and city B, respectively. We thus derive their similarity $\gamma_{A, B}$ via the Pearson correlation coefficient, and the results are shown in 
Fig.~\ref{fig:cor_city_POI_dist}. We find that NYC shows the highest similarity with PHO while the lowest similarity with SIN, implying that cities in the same country (i.e., USA) may have a higher correlation due to the similar property of culture. Besides, CAL (i.e., Canada) shows relatively higher similarity with NYC and PHO, which means that the geography property is also an important factor when measuring the correlation of cities. Although the correlation of cities can be measured from the aspect of POI distribution, the user behavioral transition pattern is a significant factor in the next POI recommendation task, we thus further explore such correlation from the angle of user sequential behaviors.

\smallskip\noindent\textbf{Correlation of Cities w.r.t Behavioral Patterns}. 
We examine the correlation of cities w.r.t. the categories of users' successive POI visits. 
In particular, given any two cities, $A^{cat} = [A_{1}^{cat}, A_{2}^{cat}...A_{\mathcal{|S|}}^{cat}]$ and $B^{cat} = [B_{1}^{cat}, B_{2}^{cat}...B_{\mathcal{|S|}}^{cat}]$ refer to the category transition distributions among $\mathcal{S}$ transition types, e.g., $A_{1}^{cat}$ denotes the ratio of transition type $FO\rightarrow SS$ within city A. Analogously, the similarity among different cities can be calculated via the Pearson correlation coefficient, shown in Fig.~\ref{fig:cor_city_behav}. Interestingly, we observe that the correlation of cities w.r.t behavioral patterns is quite different from that w.r.t POI distribution. 
Specifically, PHO and CAL still keep higher similarity, whereas NYC shows comparably lower similarity with PHO and CAL. 
To further dig out how the four cities are correlated and different over the behavioral patterns, we compare the two most correlated cities (i.e., CAL and PHO) and the two least correlated cities (i.e., NYC and SIN). 
For ease of presentation, we select the 10 most frequent category transitions for comparison as shown in Fig. 2(c-d), where the $x$-axis denotes the category transitions, e.g., $AE \rightarrow CU$ (AE2CU), and the $y-$axis shows the proportion of such a transition within a city.
We find that the more correlated cities possess consistent distributions over the frequent category transitions and \textit{vice versa}. 
The above observations depict the various correlations between cities, which inspire us to differentiate their influence when transferring knowledge from auxiliary cities to the target city.

%% file: section/framework_v1.tex
\section{The Proposed MERec}\label{Sec: TMF}
This section presents the proposed MERec, which leverages the correlation of behavioral patterns when transferring knowledge from auxiliary cities to the target city, i.e., paying more attention to more correlated knowledge.

%\subsection{Problem Formulation} 
\smallskip\noindent
\textbf{Problem Formulation.}
Each city has its unique user set $\mathcal{U}$ and POI set $\mathcal{P}$ without sharing any common users and POIs. 
For user $u$, all his check-in records, i.e., ${r = (p,c,g,t)}$, are ordered by timestamps as in~\cite{zhao2017geo}, where $p, c, g, t$ denote POI $p$, category $c$, coordinate $g$ (i.e., longitude and latitude) and timestamp $t$. 
We then split his historical records into sequences by day and obtain two types of sequences: 1) the $i$-th category sequence denoted by a set of category tuples, i.e., $C^{u,i} = \{C^{u}_{t_1}, C^{u}_{t_2}, \cdots, C^{u}_{t_n}\}$, where $C^{u}_{t_k} = (c^{u}_{t_k}, t^{u}_{k})$, and 2) the $i$-th POI sequence denoted by a set of POI tuples, i.e., $P^{u,i} = \{P^{u}_{t_1}, P^{u}_{t_2},\cdots, P^{u}_{t_n}\}$, where $P^{u}_{t_k} = (p^{u}_{t_k}, d^{u}_{t_k}, t^{u}_{k})$, and $d_{t_k}$ is the distance between successive POIs calculated by their coordinates. 
Given $C^{u,i}$, $P^{u,i}$, \textit{auxiliary cities} $\mathcal{Y}_\mathcal{A} = \{y^{(m)}_{aux}| m \in 1,2,\cdots, M\}$ and the \textit{target city} $\mathcal{Y}_\mathcal{T} = \{y_{tar}\}$, our goal is to predict user $u$'s next POI $p_{t_{n+1}}$ at time $t_{n+1}$ by transferring knowledge from the auxiliary cities to the target city.

%%%%%%%%%%%%%%%%%%%%%%%%%%%%%%%%%%%%%%%%%%%%%%%%%%%%%%%%%%%%%%%%%%%%%%%%%%%%%%%%%
%%%%%%%%%%%%%%%%%%%%%%%%%%%%%%%%%%%%%%%%%%%%%%%%%%%%%%%%%%%%%%%%%%%%%%%%%%%%%%%%%
\begin{figure*}[t]
\centering
\includegraphics[width=1\textwidth]{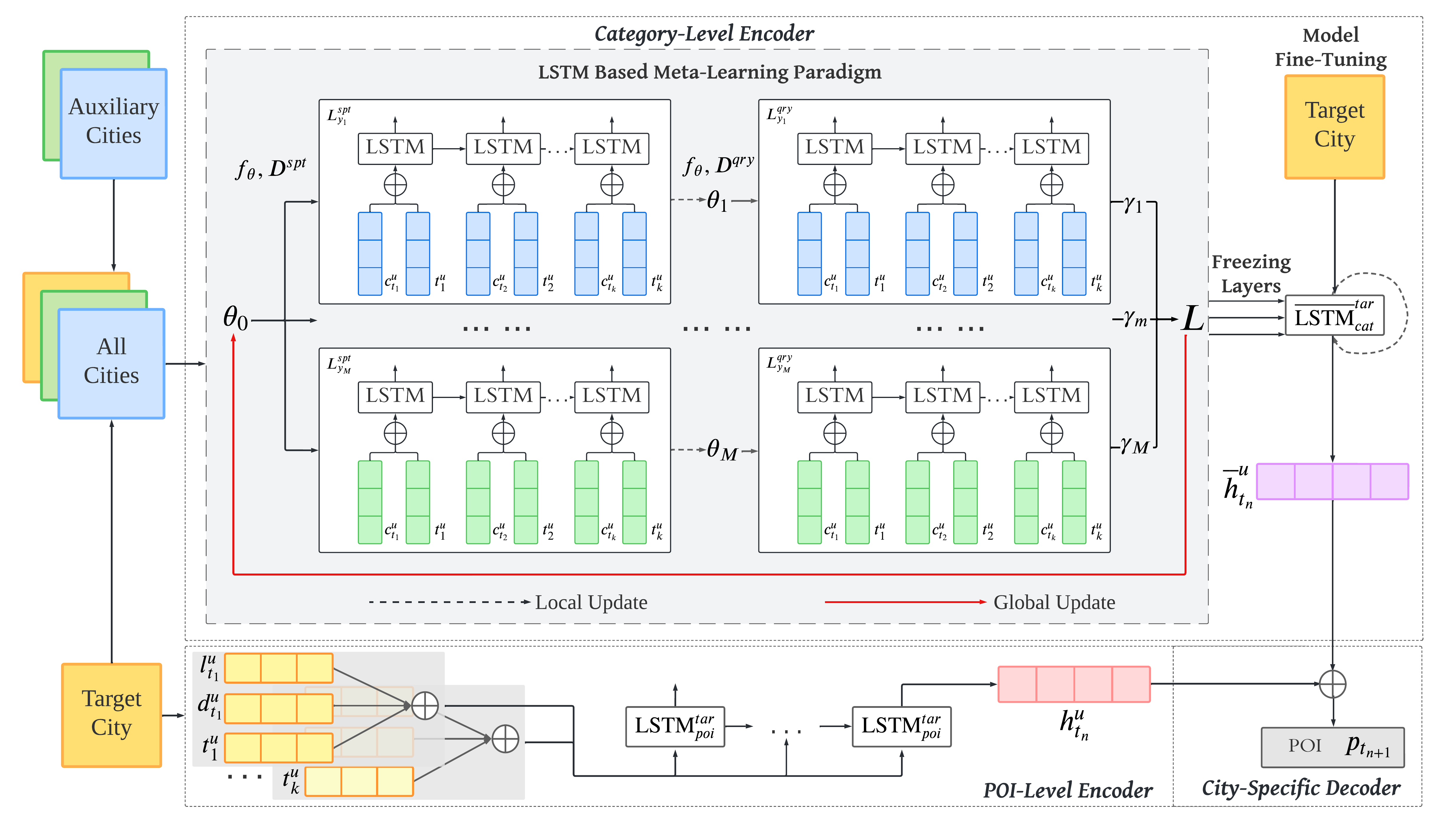}
\vspace{-0.3in}
\caption{The overall framework of our proposed MERec.}\label{fig:2}
\vspace{-0.25in}
\end{figure*}

%\subsection{The MERec Framework}
\smallskip\noindent
\textbf{Overview of MERec.}
The overview of MERec is outlined in Fig.~\ref{fig:2}, mainly composed of a \textit{two-channel encoder} (i.e., category- and POI-level encoders)  with the embedding layer and a \textit{city-specific decoder}. 
In particular, the category-level encoder exploits meta-learning to capture the common user check-in transition patterns at the category level in each city by holding the principle of ``paying more attention to more correlated knowledge''. 
The goal of the POI-level encoder is to learn the accurate POI transition patterns in the target city. Lastly, the city-specific decoder performs the next POI predictions by concatenating the hidden states of the above two encoders.

\smallskip\noindent\textbf{Embedding Layer.} 
It maps each check-in record into an embedding vector. Specifically, in the category-level encoder, the embedding of a category tuple $\mathbf{e}^{C} \in \mathbb{R}^{2d}$ is the concatenation of the category embedding $\mathbf{e}^{c}\in \mathbb{R}^{d}$ and time embedding $\mathbf{e}^{t}\in \mathbb{R}^{d}$; thus the embedding of a category sequence $C^{u,i}$ is formed as $\mathbf{E}_{C^{u,i}}=[\mathbf{e}^{C}_{t_{1}}, \mathbf{e}^{C}_{t_{2}},\cdots, \mathbf{e}^{C}_{t_{n}}]$. 
Analogously, in the POI-level encoder, the embedding  of a POI sequence is denoted by $\mathbf{E}_{P^{u,i}}=[\mathbf{e}^{P}_{t_{1}}, \mathbf{e}^{P}_{t_{2}},\cdots, \mathbf{e}^{P}_{t_{n}}]$, where $\mathbf{e}^{P}$ is the embedding of POI tuple represented by the concatenation of POI embedding $\mathbf{e}^{p}\in \mathbb{R}^{d}$, distance embedding $\mathbf{e}^{dist}\in \mathbb{R}^{d}$ and time embedding $\mathbf{e}^{t}\in \mathbb{R}^{d}$.

\smallskip\noindent\textbf{Cateogry-level Encoder.} To distil knowledge from auxiliary cities and employ category-level user behavioral patterns, we extend model-agnostic meta-learning (MAML)~\cite{finn2017model} with LSTM as the framework for the meta-learning update. 
In particular, we devise a \textit{correlation strategy} that can transfer knowledge based on the correlation of user behavioral patterns among cities. 
Meanwhile, \textit{freezing layers and model fine-tuning} are exploited to obtain a generic model while better adapting to the data of the target city.

\smallskip\noindent\textit{\underline{Meta-learning Setup.}} 
Following~\cite{chen2021curriculum}, the recommendation within each city, including the auxiliary and target cities, can be viewed as a single task (with its own dataset $\mathcal{D}$) in a meta-learning paradigm. Thus, the check-in sequences of auxiliary cities $\mathcal{Y}_{A}$ are denoted as $\mathbb{D}_{meta}^{(aux)}$, 
and the check-in sequences of target city $\mathcal{Y}_{T}$ are divided as training set $\mathbb{D}_{train}^{(tar)}$ and test set $\mathbb{D}_{test}^{(tar)}$.  
We treat each city $y_m$ as a meta-learning task, where each task has support set $\mathcal{D}^{spt}_{y_m}$ for training and a query set $\mathcal{D}^{qry}_{y_m}$ for testing. 
Finally, our goal is to leverage the data from both auxiliary cities and the target city, i.e., $\mathbb{D}_{train}=\mathbb{D}_{meta
}^{(aux)}\cup\mathbb{D}_{train}^{(tar)}$, to learn a meta-learner $F_w$, where $w$ is its parameters. Accordingly, given the support sets, $F_w$ predicts the parameters $\theta$ of recommender $f_\theta$ to minimize the recommendation loss on the query sets across all cities as follows,
\begin{equation}
\footnotesize
w^* = \underset{w}{arg\,min} \sum_{y_m \in \{\mathcal{Y}_\mathcal{A}\cup \mathcal{Y}_\mathcal{T}\}} \mathcal{L}(f_\theta,\mathcal{D}^{qry}_{y_m}|\mathbb{D}_{train},\mathcal{D}^{spt}_{y_m}), \;
s.t. \; \theta = F_w(\mathcal{D}^{spt}_{y_m}|\mathbb{D}_{train
}).
\end{equation}

Specifically, each iteration of MAML includes \textcolor{black}{local update and global update on the sampled task batch, where the first phase updates $\theta$ locally on $\mathcal{D}^{spt}$ of each task, and the second phase globally updates $\theta$ by gradient descent to minimize the sum of loss on $\mathcal{D}^{qry}$ of all tasks.}

\begin{itemize}[leftmargin=*]
    \item \textit{Local update:} we first sample a batch of cities, and then randomly sample $N$ category sequences $\mathcal{D}^{spt}_{y_m}$ and $\mathcal{D}^{qry}_{y_m}$ for each sampled city. Thus, we calculate the training loss on  $\mathcal{D}^{spt}_{y_m}$ and locally update %$\theta^{(cat)}$ 
    $\theta$ by one step:
    \begin{equation} \label{equ1}
    \footnotesize
     \theta_{y_m}^{\prime}=\theta-\alpha \nabla_{\theta} \mathcal{L}_{y_m} (f_\theta,\mathcal{D}^{spt}_{y_m}),
    \end{equation}
    where $\mathcal{L}$ is the cross-entropy loss; $\alpha$ is the local learning rate, and \textcolor{black}{$\theta_{y_m}^{\prime}$} is the locally updated parameters of recommender for each city. 
    \item \textit{Global update:} we calculate the testing loss on each $\mathcal{D}^{qry}_{y_m}$ with the corresponding $ \theta_{y_m}^{\prime}$ and then update the initialization $\theta$ by one gradient step on the sum of testing losses across all cities, where $\beta$ is the global learning rate.
    \begin{equation}\label{equ2}
    \footnotesize
     \theta =\theta-\beta \nabla_{\theta} \sum\nolimits_{y_m \in \{\mathcal{Y}_\mathcal{A}\cup \mathcal{Y}_\mathcal{T}\}} \mathcal{L}_{y_m} (f_{\theta'_{y_m}},\mathcal{D}^{qry}_{y_m}).
    \end{equation}
\end{itemize}

\noindent\textit{\underline{Correlation Strategy.}} 
From the data analysis in Section 3, we observe that there exist various correlations w.r.t different aspects among different cities. 
Directly transferring user check-in behaviors from auxiliary cities to the target city may introduce noise thus hurting the recommendation performance. 
By holding the principle of ``paying more attention to more correlated knowledge'', 
we further consider the \textit{correlation of behavioral patterns at category level} in different cities when conducting the global update. 
To be specific, we obtain the city-level correlation (e.g., $\gamma_{cor}$) based on behavioral patterns, and then attentively adapt the gradient across cities by employing their correlations. 
In other words, if the auxiliary city is more correlated to the target city, we adapt the gradient so that it updates faster in that direction. 
Therefore, Eq.(\ref{equ1}) is reformulated as:
\begin{equation}\label{equ3}
\footnotesize
 \theta_{y_m}^{\prime}=\theta-\alpha \nabla_{\theta} [\mathcal{L}_{y_m} (f_\theta,\mathcal{D}^{spt}_{y_m}) \times\gamma_{cor}].
\end{equation}

\noindent\textit{\underline{Freezing Layers and Model Fine-Tuning.}} Inspired by~\cite{yosinski2014transferable}, the network with freezing layers and fine-tuning is generalized better than the one trained directly on the target dataset. Therefore, after obtaining the well-trained category-level encoder for the target city (i.e., $LSTM_{cat}^{tar}$) by the meta-learning paradigm, we further consider fine-tuning it. In doing this, we can deliver a network that not only accommodates knowledge distilled from the auxiliary cities but also better adapts to the target city. Specifically, assuming $LSTM_{cat}^{tar}$ contains $L$ layers, we freeze its first $l$ $(1\leq l \leq L)$  layers, while adding $n$ layers after the $l$ layers. The newly constructed model is denoted by $\overline{LSTM}_{cat}^{tar}$, which is further fine-tuned via category sequences from the target city, i.e., $\mathbb{D}_{train}^{(tar)}$. As such, the freezing-layers help generate a network that can better balance parameters between auxiliary cities and the target city after the fine-tuning.
Accordingly, the hidden state $\boldsymbol{\overline{h}}^u_{t_k}$ of category at $t_k$ is given by,
\begin{equation}\label{equ4}
\footnotesize
\boldsymbol{\overline{h}}^u_{t_k} = \overline{LSTM}_{cat}^{tar}(\bold{e}_{t_k}^{C}, \bold{\overline{h}}_{t_{k-1}}^u).
\end{equation}
%

%\subsubsection{POI-level Encoder}
\smallskip\noindent\textbf{POI-level Encoder.} 
It aims to model users' sequential check-in behaviors and the spatio-temporal context in the target city by using the LSTM model. 
As illustrated in the \textit{Embedding Layer}, the embedding of a POI sequence is represented by $\mathbf{E}_{P^{u,i}}=[\mathbf{e}^{P}_{t_{1}}, \mathbf{e}^{P}_{t_{2}},\cdots, \mathbf{e}^{P}_{t_{n}}]$, where each embedding $\mathbf{e}^{P}_{t_{k}}$ is feed into the $LSTM_{poi}^{tar}$ to infer the hidden state 
 $\boldsymbol{{h}^u_{t_k}}$ of POI check-in at $t_k$, given by,
\begin{equation}\label{equ5}
\footnotesize
\boldsymbol{{h}^u_{t_k}} = LSTM_{poi}^{tar}(\mathbf{e}^{P}_{t_{k}}, \boldsymbol{h}^u_{t_{k-1}}).
\end{equation}

\smallskip\noindent\textbf{City-specific Decoder.} 
The city-specific decoder aims to perform the next POI prediction based on the last hidden states learned from the two-channel encoder (i.e., $\overline{\mathbf{h}}^u_{t_n}, \mathbf{h}^u_{t_n}$).  Accordingly, the probability distribution on all candidate POIs is calculated by the softmax function, given by,
\begin{equation}\label{equ:predict}
\footnotesize
\boldsymbol{\hat{y}} = softmax(f(\overline{\mathbf{h}}^u_{t_n}; \mathbf{h}^u_{t_n})),
\end{equation}
where $f$ is a fully connected layer to transform $(\overline{\mathbf{h}}^u_{t_n}; \mathbf{h}^u_{t_n})$ into a $|\mathcal{P}|$-dimensional vector; and $|\mathcal{P}|$ is the number of POIs in the target city. Hence, the objective function for the next POI recommendation is defined by: 
\begin{equation}\label{equ:loss}
\footnotesize
\mathcal{J} =  -\sum\nolimits_{i=1}^{|\mathcal{P}|} \boldsymbol{y}[i]\cdot log \space (\hat{\boldsymbol{y}}[i]),
\end{equation}
where $\boldsymbol{y}$ is a one-hot embedding of the ground-truth POI. 
Algo.~\ref{alg:1} shows the training process of MERec, consisting of meta training (lines 3-9), freezing layers and model fine-tuning (lines 10-12), as well as next POI prediction (lines 13-14). 

\input{tables/algorithm.tex}

%% file: tables/algorithm.tex
\begin{algorithm}[t]
\footnotesize
\caption{The training process of MERec}\label{alg:1}
\LinesNumbered 
\KwIn{$\mathbb{D}_{train},\mathcal{Y}_\mathcal{A}, \mathcal{Y}_\mathcal{T}, \alpha,\beta, Iter, N, l, n$ 
%the number of shots $N$ and iterations of meta-learning $M$
}     
\KwOut{A list of recommended next POIs}
Randomly initialize parameters $\theta$\;
%$\theta = \theta^{(cat)} \cup\theta^{(poi)}$\;
%
Calculate the correlation of behavioral patterns at category level\;
\For{$(iter=1; iter\leq Iter; iter++)$}
{
    \For{each city $y_m \in \{\mathcal{Y}_\mathcal{A} \cup \mathcal{Y}_\mathcal{T}\}$}
    {
        Sample $N$ category sequences from $\mathbb{D}_{y_m}^{spt}$ as the adapt\_batch\;
        Evaluate: $\nabla_{\theta}\mathcal{L}_{y_m}(f_\theta,\mathcal{D}^{spt}_{y_m})$ using the adapt\_batch\;
        Calculate the gradient update of $\theta'_{y_m}$ by Eq.(\ref{equ3});  \tcp{local update} 
        Sample $N$ category sequences from $\mathbb{D}_{y_m}^{qry}$ as the eval\_batch\;
    }
    Update $\theta$ using eval\_batch by Eq.(\ref{equ2}); \tcp{global update}
}
Freeze the first $l$ layers and add $n$ layers as the new $\overline{LSTM}_{cat}^{tar}$ model\; 
Fine-tune $\overline{LSTM}_{cat}^{tar}$ via the training category sequences of the target city\;
Get the last hidden states of the two-channel encoder shown in Eqs. (\ref{equ4}-\ref{equ5})\;
Predict the next possible POI via Eq.(\ref{equ:predict})\;
Calculate the prediction loss for each check-in record via Eq.(\ref{equ:loss})\;
\end{algorithm}

%% file: section/experiments_v1.tex
\section{Experiments and Results}\label{Sec: ER}
We conduct experiments to answer three research questions: \textbf{(RQ1)} does MERec outperform state-of-the-art baselines? \textbf{(RQ2)} how do different components of MERec affect its performance? \textbf{(RQ3)} how do essential hyper-parameters affect MERec? The code is available at \url{https://github.com/oli-wang/MERec}.

%\subsection{Experimental Setup}
\smallskip\noindent\textbf{Datasets and Evaluation Metrics.} The four datasets shown in Table~\ref{tab1} are used in our experiment, where we take one of the cities as the target city and the rest as auxiliary cities each time. 
Following~\cite{huang2019attention}, 
we chronologically divide the dataset of the target city into training, validation, and test sets with a ratio of 8:1:1.
Note that we remove users and POIs with less than five and three check-ins, respectively. 
Two commonly-used metrics, i.e., $HR@K$ and $NDCG@K$ are adopted by following~\cite{chen2021curriculum},
where the former measures whether the ground-truth POI can be found in the top-$K$ recommendation list, and the latter measures the ranking quality of the ground-truth POI in the recommendation list.

\smallskip\noindent\textbf{Compared Baselines.} We compare the MERec with seven state-of-the-art approaches. (1) \texttt{MostPop} recommends the next POI based on the popularity of POIs; (2) \texttt{BPRMF} is a matrix factorization method optimized via Bayesian personalized ranking; (3) \texttt{NeuMF}~\cite{he2017neural} generalizes the matrix factorization by employing a multi-layer perceptron to model the user-item interactions; (4) \texttt{ATST-LSTM}~\cite{huang2019attention} is an attention-based LSTM method by considering spatio-temporal contextual information; (5) \texttt{iMTL}~\cite{zhang2021interactive} is a multi-task learning framework for next POI recommendation, which consists of a two-channel encoder and a task-specific decoder; (6) \texttt{MAML}~\cite{finn2017model} is a model-agnostic meta-learning for few-shot learning tasks; (7) \texttt{CHAML}~\cite{chen2021curriculum} is a meta-learning based framework for next POI recommendation, which considers both city- and user-level hardness during meta training.

\smallskip\noindent\textbf{Hyper-parameter Settings.} 
The optimal hyper-parameter settings for all methods are empirically found out based on the performance on the validation set. Specifically, the embedding size is searched from $\{32, 64, 128, 256\}$. For baselines (2-5), the learning rate is selected from $\{0.1, 0.05, 0.01, 0.005, 0.001, 0.0001\}$, and the batch size is set as 256. For meta-learning based baselines (6-7) and MERec, the learning rates $\alpha, \beta$ are searched from \{0.5, 0.1, 0.01, 0.001, 0.0001\}; and the batch size is set as 256 for a fair comparison. For MERec, the number of freezing layers $l$ is searched in the range of $[1, 4]$ stepped by one, where the best setting is 3 for all cities; and $Iter=500, N=32, n=2$ across all cities.   

\input{tables/table_results_all}

%\subsection{Performance Comparison (RQ1)} 
\smallskip\noindent\textbf{Performance Comparison (RQ1).}
The results are presented in Table~\ref{table:comparison-results-all}. 
Across the four datasets, the traditional methods (MostPop, BPRMF) generally perform worse than deep learning methods (NeuMF, ATST-LSTM, iMTL) demonstrating the efficacy of neural networks on more accurate recommendation. 
RNN based methods (ATST-LSTM, iMTL) outperform NeuMF, which indicates the capability of RNN on modeling the sequential dependency. 
iMTL defeats ATST-LSTM, as it leverages multi-task learning (MTL) framework to jointly learn user preference on both categories and POIs, 
exhibiting the superiority of MTL on better next POI recommendation.
Meta-learning based methods (MAML, CHAML, MERec) bring further enhancement compared with other methods, showcasing the efficacy of knowledge transfer in alleviating the data sparsity issue. Overall, our MERec consistently achieves the best performance across all the datasets, with an average lift of 6.3\% and 6.53\% w.r.t. HR and NDCG, respectively. This helps confirm the benefits of (1) leveraging check-ins of auxiliary cities to augment the target city, and (2) paying more attention to more correlated knowledge when transferring knowledge from auxiliary cities.

\begin{figure*}[t]
\centering 
\includegraphics[width=1\textwidth]{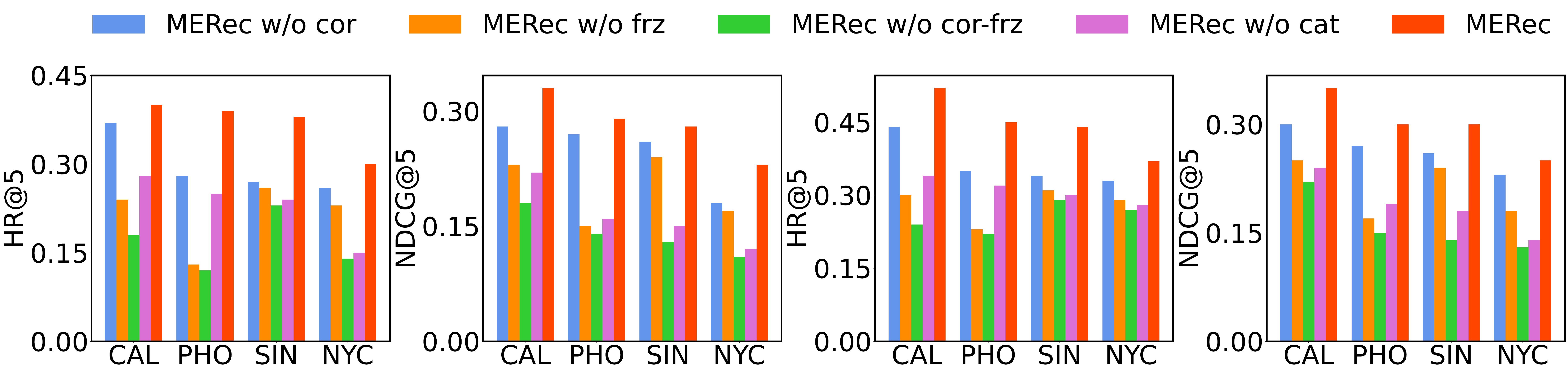} 
\vspace{-0.2in}
\caption{Performance comparison for variants of MERec on the four datasets.}
\label{Fig:ablation} 
%\vspace{-0.1in}
\end{figure*}

%\subsection{Ablation Study (RQ2)}
\smallskip\noindent\textbf{Ablation Study (RQ2)}.
To check the impacts of various components in MERec, four variants are compared. 
(1) MERec$_{w/o \ cor}$ removes the correlation strategy from the meta-learner; (2) MERec$_{w/o \ frz}$ removes the freezing layers and fine-tuning from the category-level encoder; (3) MERec$_{w/o \ cor-frz}$ removes both correlation strategy, freezing layers and fine-tuning; and (4) MERec$_{w/o \ cat}$ removes the category-level encoder, but only retains the POI-level encoder. The results are shown in
Fig.~\ref{Fig:ablation}.
We note that MERec$_{w/o \ cor-frz}$ performs worse than both MERec$_{w/o \ cor}$ and MERec$_{w/o \ frz}$, suggesting that both the correlation strategy, freezing layers, and fine-tuning operation indeed improve the recommendation performance. 
Generally, the performance decrease of MERec$_{w/o \ frz}$ far exceeds that of MERec$_{w/o \ cor}$, implying that the freezing layers and fine-tuning operation play more important roles than the correlation strategy. 
Besides, MERec$_{w/o \ cat}$ underperforms MERec, which helps verify the advantages of both the meta-learning paradigm with auxiliary check-ins and the correlation strategy. 

%Figure of Sensitivity on local-update steps of Meta-learning
\begin{figure}[t]
\centering
\subfigure[]{
\begin{minipage}[t]{0.22\linewidth}
\centering
\includegraphics[width=1.15in]{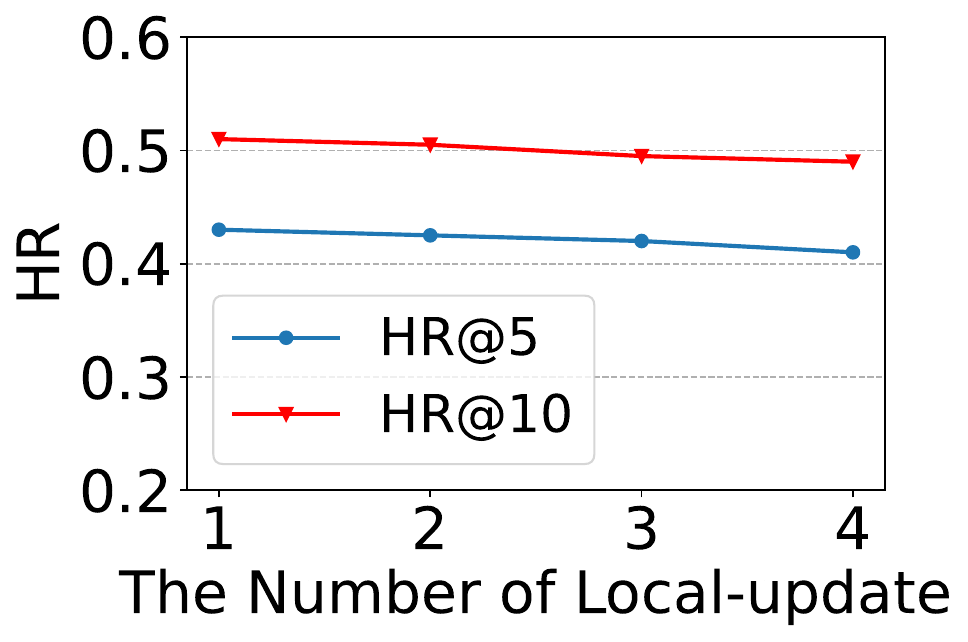}
\end{minipage}
}
\subfigure[]{
\begin{minipage}[t]{0.22\linewidth}
\centering
\includegraphics[width=1.2in]{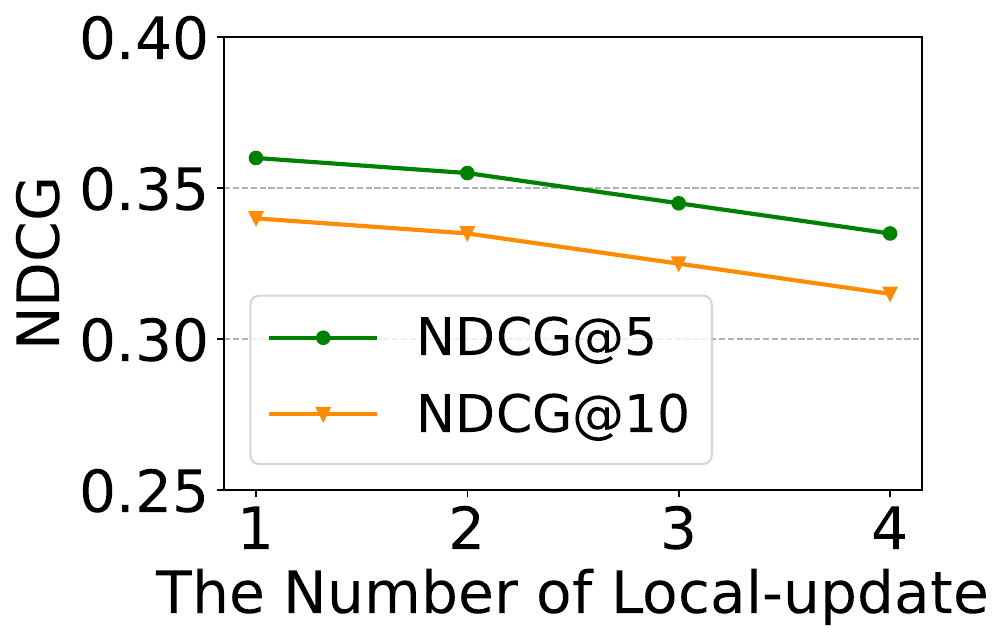}
%\caption{fig2}
\end{minipage}
}
\subfigure[]{
\begin{minipage}[t]{0.22\linewidth}
\centering
\includegraphics[width=1.2in]{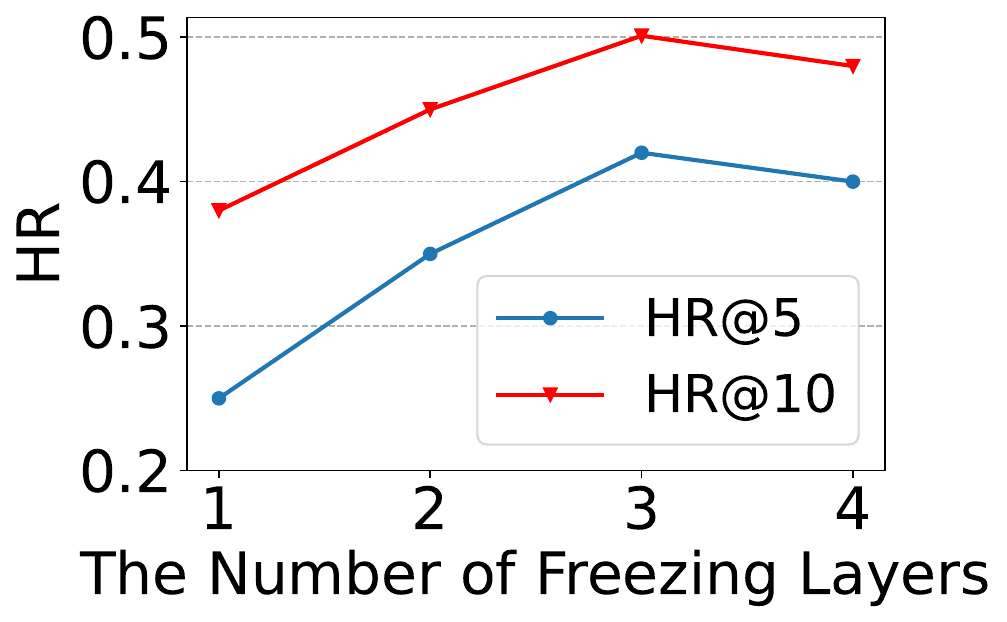}
%\caption{fig2}
\end{minipage}
}
\subfigure[]{
\begin{minipage}[t]{0.22\linewidth}
\centering
\includegraphics[width=1.2in]{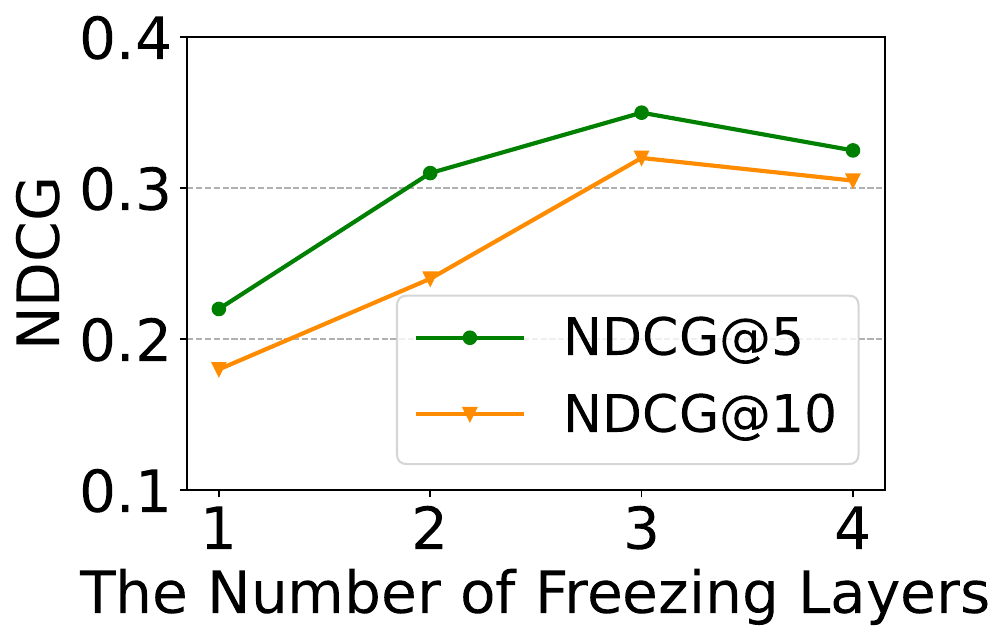}
%\caption{fig2}
\end{minipage}
}
\centering
\vspace{-0.2in}
\caption{Parameter sensitivity analysis on CAL.}\label{fig:hyper_para}
\vspace{-0.25in}
\end{figure}

%\subsection{Parameter Sensitivity Analysis}
\smallskip\noindent\textbf{Parameter Sensitivity Analysis (RQ3)}.
We study the influence of two essential hyper-parameters, i.e., the number of local-update steps in Eq.(\ref{equ1}) and the number of freezing layers. 
Fig.~\ref{fig:hyper_para} only reports the results on the CAL dataset and similar trends can be observed on the rest three datasets. 
Figs.~\ref{fig:hyper_para} (a-b) depict the model performance w.r.t. the number of local-update steps. We empirically find out that updating only one step is sufficient to obtain better recommendation accuracy, which also increases the model efficiency.  
Figs.~\ref{fig:hyper_para} (c-d) display the influence of the number of layers frozen on the model performance. As observed, with the layer increasing, the performance first goes up and then drops slightly. The best setting for the number of freezing layers is 3 on the four datasets.

%% file: tables/table_results_all.tex
\begin{table}[t]
\renewcommand{\arraystretch}{1.2}
\centering
\scriptsize
\addtolength{\tabcolsep}{1pt}
\caption{Comparative results of all approaches on the four datasets, where `H' refers to `HR' and `N' means `NDCG'; the best results are highlighted in bold; the runner up is underlined; and the column `Improve' indicates the improvements achieved by MERec relative to the runner up.}\label{table:comparison-results-all}
\begin{tabular}{cl|cc|ccc|ccc|r}
\toprule
& & \multicolumn{2}{c|}{\textbf{Traditional}} 
& \multicolumn{3}{c|}{\textbf{Deep Learning}}     
& \multicolumn{3}{c|}{\textbf{Meta Learning}}
& \multirow{2}{*}{\textit{Improve}} \\
&  & MostPop & BPRMF & NeuMF  & ASTA-LSTM & iMTL   & MAML & CHAML   & MERec &  \\ \midrule
\multirow{4}{*}{\rotatebox[origin=c]{90}{\textbf{CAL}}}   
&{H@5}  & 0.0988          & 0.1304         & 0.1431 & 0.2924    & 0.2652 & 0.3987  & \underline{0.3995} & \textbf{0.4274} & 6.98\%                            \\
%%%%%%
&{H@10} & 0.1547          & 0.2349         & 0.2368 & 0.3705    & 0.3184 & 0.4618       & \underline{0.4777} & \textbf{0.5054} & 5.80\%                            \\
%%%%%%%
& {N@5}  & 0.0632          & 0.0928         & 0.0989 & 0.2134    & 0.1857 & \underline{0.3178} & 0.3093       & \textbf{0.3378} & 6.29\%                            \\
%%%%%%%%
&{N@10} & 0.0814          & 0.1672         & 0.1669 & 0.2383    & 0.2299 & \underline{0.3362} & 0.3315       & \textbf{0.3564} & 6.01\% \\ \hline
%%%%%%%%%
\multirow{4}{*}{\rotatebox[origin=c]{90}{\textbf{PHO}}}   &{H@5}  & 0.0682          & 0.1093         & 0.1316 & 0.2366    & 0.2410 & 0.3549       & \underline{0.3660} & \textbf{0.3928} & 7.32\%                            \\
%%%%%%%%%
&{H@10} & 0.1068          & 0.1584         & 0.1852 & 0.3125    & 0.3370 & \underline{0.4508} & 0.4419       & \textbf{0.4531} & 0.51\%                            \\
%%%%%%%
& {N@5}  & 0.0419          & 0.0688         & 0.0869 & 0.1635    & 0.1753 & 0.2633       & \underline{0.2648} & \textbf{0.2796} & 5.59\%                            \\
%%%%%%%%%
& {N@10} & 0.0547          & 0.0848         & 0.1042 & 0.1883    & 0.2065 & \underline{0.2949} & 0.2891       & \textbf{0.2993} & 1.49\%                            \\ \hline
\multirow{4}{*}{\rotatebox[origin=c]{90}{\textbf{SIN}}} & {H@5}  & 0.0365   & 0.0848         & 0.1004 & 0.2165    & 0.2388 & 0.2991       & \underline{0.3571} & \textbf{0.3784} & 5.96\%  \\
& {H@10} & 0.0635          & 0.1450         & 0.1696 & 0.2879    & 0.3080 & 0.3816       & \underline{0.4486} & \textbf{0.4557} & 1.58\%                            \\
                           & {N@5}  & 0.0231          & 0.0452         & 0.0697 & 0.1532    & 0.1696 & 0.2188       & \underline{0.2650} & \textbf{0.2749} & 3.73\%                            \\
                           & {N@10} & 0.0318          & 0.0648         & 0.0925 & 0.1760    & 0.1922 & 0.2451       & \underline{0.2981} & \textbf{0.3015} & 1.14\%                            \\ \hline
\multirow{4}{*}{\rotatebox[origin=c]{90}{\textbf{NYC}}}  & {H@5}  & 0.0214          & 0.0558         & 0.0959 & 0.1763    & 0.2187 & 0.2456     &\underline{0.2745} & \textbf{0.2991} & 8.96\%                            \\
                           & {H@10} & 0.0336          & 0.0994         & 0.1495 & 0.2455    & 0.2879 & 0.3373       & \underline{0.3526} & \textbf{0.3995} & 13.30\%                            \\
                           & {N@5}  & 0.0134          & 0.0265         & 0.0595 & 0.1257    & 0.1484 & 0.1652       & \underline{0.1865} & \textbf{0.2107} & 12.98\%                           \\
                           & {N@10} & 0.0173          & 0.0237         & 0.0770 & 0.1485    & 0.1705 & 0.2072       & \underline{0.2118} & \textbf{0.2436} & 15.01\%                           \\ \bottomrule
\end{tabular}
\vspace{-0.15in}
\end{table}

%% file: section/conclusion.tex
\section{Conclusion}
In this paper, we propose a Meta-learning Recommendation (MERec) framework for the next POI recommendation by leveraging check-ins from auxiliary cities to augment the target city, and holding the principle of ``paying more attention to more correlated knowledge''. 
In particular, we devise a two-channel encoder to capture the transition patterns of categories and POIs, whereby a city-correlation based strategy is devised to attentively capture common knowledge (i.e., patterns) from auxiliary cities via the meta-learning paradigm. The city-specific decoder then concatenates the latent representations of the two-channel encoder to perform the next POI prediction for the target city. Extensive experiments on four real-world datasets demonstrate the superiority of our proposed MERec.